# Amplifying the Impact of Open Access: Wikipedia and the Diffusion of Science

**Misha Teplitskiy**
Dept. of Sociology and
KnowledgeLab
University of Chicago
misha@uchicago.edu
(773) 834-4787
5735 South Ellis Avenue
Chicago, Illinois 60637

**Grace Lu**
KnowledgeLab
University of Chicago
gracelu@uchicago.edu
(773) 834-4787
5735 South Ellis Avenue
Chicago, Illinois 60637

**Eamon Duede**
Computation Institute and
KnowledgeLab
University of Chicago
eduede@uchicago.edu
(773) 834-4787
5735 South Ellis Avenue
Chicago, Illinois 60637

## Abstract
With the rise of Wikipedia as a first-stop source for scientific knowledge, it is important to compare its representation of that knowledge to that of the academic literature. Here we identify the 250 most heavily used journals in each of 26 research fields (4,721 journals, 19.4M articles in total) indexed by the *Scopus* database, and test whether topic, academic status, and accessibility make articles from these journals more or less likely to be referenced on Wikipedia. We find that a journal's academic status (impact factor) and accessibility (open access policy) both strongly increase the probability of its being referenced on Wikipedia. Controlling for field and impact factor, the odds that an open access journal is referenced on the English Wikipedia are 47% higher compared to paywall journals. One of the implications of this study is that a major consequence of open access policies is to significantly amplify the diffusion of science, through an intermediary like Wikipedia, to a broad audience.

## Introduction
Wikipedia, one of the most visited websites in the world[1], has become a destination for information of all kinds, including information about science (Heilman & West, 2015; Laurent & Vickers, 2009; Okoli, Mehdi, Mesgari, Nielsen, & Lanamäki, 2014; Spoerri, 2007). Given that so many people rely on Wikipedia for scientific information, it is important to ask whether and to what extent Wikipedia's coverage of science is a balanced, high quality representation of the knowledge within the academic literature. One approach to asking this question involves looking at references used in Wikipedia articles. Wikipe-dia requires all claims to be substantiated by reliable references[2], but what, in practice, are "reliable references?"

An intuitive approach is to examine whether the sources Wikipedia editors use correspond to the sources scientists value most. In particular, within the scientific literature, a journal's status is often associated, albeit problematically (Seglen, 1997), with its impact factor. If status within the academic literature is taken as a "gold standard," Wikipedia's failure to cite high impact journals of certain fields would constitute a failure of coverage (Samoilenko & Yasseri, 2014), while a high correspondence between journals' impact factors and citations in Wikipedia would indicate that Wikipedia does indeed use reputable sources (P. Evans & Krauthammer, 2011; Nielsen, 2007; Shuai, Jiang, Liu, & Bollen, 2013).

Yet high impact journals often require expensive subscriptions (Björk & Solomon, 2012). The costs are, in fact, so prohibitive that even Harvard University has urged its faculty to "resign from publications that keep articles behind paywalls" because the library "can no longer afford the price hikes imposed by many large journal publishers" (Sample, 2012). Consequently, much of the discussion of open access focuses on the consequences of open access for the scientific community (Van Noorden, 2013). A lively debate has arisen on the impact of open access on the scientific literature, with some studies showing a citation advantage (Eysenbach, 2006a, 2006b; Gargouri et al., 2010; "The Open Access Citation Advantage Service") while other find none (Davis, Lewenstein, Simon, Booth, & Connolly, 2008; Davis, 2011; Gaulé & Maystre, 2011; Moed, 2007).

Apart from a rather unclear impact on the scientific literature, open access journals may have a tremendous impact on the diffusion of scientific knowledge *beyond* this literature. To date, this potential of open access policies has been a matter chiefly of speculation (Heilman & West, 2015; Trench, 2008). Previous research has found that open access articles are downloaded from publishers' websites more often and by more people than closed access articles (Davis, 2010, 2011), but it is currently unclear by whom, and to what extent open access affects the use of science by the *general public* (Davis & Walters, 2011). We hypothesize that Wikipedia, with more than 8.5 million page views *per hour*[3], diffuses scientific knowledge to unprecedented distances and that diffusion of science through it may relate to accessibility in two ways. By referencing findings from paywall journals, Wikipedia distills and diffuses these findings to the general public. On the other hand, Wikipedia editors may be unable to access expensive paywall journals[4], and consequently reference the easily accessible articles instead. For example, Luyt and Tan's (Luyt & Tan, 2010) study found accessibility to drive the selection of references in a sample of Wikipedia's history articles. In this case Wikipedia "amplifies" open access science by broadcasting its (already freely accessible) findings to millions. This "amplifier" effect may thus constitute one of the chief effects of open access.

### *Correspondence between academic and Wikipedia statuses*

This article tests both the distillation and amplifier hypotheses by evaluating which references Wikipedia editors around the world use and do not use. In particular we study the correspondence between journals' status within the scientific community (impact factor) and their accessibility (open access policy) with their status within Wikipedia (percent of a journal's articles referenced in Wikipedia). It is important to note that an observed correspondence may be evinced by a variety of mechanisms besides the aforementioned accessibility. First, the status ordering of academic journals as measured with impact factors may have only a tenuous relationship with the importance and notability – considerations of special relevance to Wikipedia[5] – of the published research. Citations, and therefore impact factors, are in part a function of the research field (Seglen, 1997), and may be affected by factors as circumstantial as whether a paper's title contains a colon (Jamali & Nikzad, 2011; Seglen, 1997). Second, the academic status ordering results from the objectives of millions of scientists and institutions, and may be irrelevant to the unique objectives of Wikipedia. Wikipedia's key objective is to serve as an encyclopedia, not a medium through which scientists communicate original research[6]. Relative to the decentralization of the scientific literature, Wikipedia is governed by explicit, if flexible, policies and a hierarchical power structure (Butler, Joyce, & Pike, 2008; Shaw & Hill, 2014). Apart from a remark that review papers serve Wikipedia's objectives better than primary research articles, Wikipedia's referencing policies generally pass no judgment over which items within the scientific literature constitute "the best" evidence in support of a claim[7]. Wikipedia's objectives and explicit, centrally accessible, policies differ from the decentralized decisions that produce status orderings within the scientific literature and do not imply that the two status orderings should correspond. Indeed, if editors are not scientists themselves they need not even be aware that journal impact factors exist[8]. On the other hand, despite the well-worn caveats, prestigious, high-impact journals may publish findings that are more important to both academics and Wikipedia's audience. In fact, a Wikipedia editor's *expectation* that the truly important research resides within high-impact journals may be enough to predispose them to want reference such journals. Second, little is known about editors of science-related articles (West, Weber, & Castillo, 2012); they may be professional scientists with access to these high-impact journals, resulting in both the *motivation* and *opportunity* to reference them.

## Previous research

### *Wikipedia references and academic status*

The first large-scale study of Wikipedia's scientific references was performed by Finn Arup Nielsen (Nielsen, 2007). Nielsen found that the number of Wikipedia references to the top 160 journals, extracted from the *cite-journal* citation templates, correlated modestly with the journal's *Journal Citation Reports* impact factor. This implication that Wikipedia preferentially cites high impact journals is delicate in part because the data used in the study included only a subset of journals with references that appear in Wikipedia, not journals that were and *were not* referenced. It is possible, albeit unlikely, that an even larger number of prestigious journals, made invisible by the methodology, are never referenced on Wik-ipedia at all, weakening the correlation to an unknown degree, or that the referenced journals are simply those that publish the most articles (see Nielsen 2007: Fig. 1). Shuai, Jiang, Liu, and Bollen (2013) also found modest correlations when they investigated a possible correspondence between the academic rank of computer science papers, authors, and topics and their Wikipedia rank.

The altmetrics movement has also explored Wikipedia as non-academic venue on which academic literature makes an impact (ALM, Fenner, & Lin, 2014; "altmetrics," ; Priem, 2015). Evans and Krauthammer (P. Evans & Krauthammer, 2011) examined the use of Wikipedia as an alternative measure of the scholarly impact of biomedical research. The authors correlated scholarly metrics of biomedical articles, journals, and topics with Wikipedia citations and, in contrast to other studies, included in some of their analyses a random sample of journals never referenced on Wikipedia. The authors also recorded a journal's open access policy but, unfortunately, do not appear to have used this information in analyses.

### *Open access and the Web*

The rather voluminous literature on open access has focused primarily on effects on the academic literature[9]. There is some debate on the size and direction of open access effects. Some evidence demonstrates that open access articles gain a citation advantage (Eysenbach, 2006a, 2006b; Gargouri et al., 2010; "The Open Access Citation Advantage Service"), while other evidence shows no such effect (Davis et al., 2008; Davis, 2011; Gaulé & Maystre, 2011; Moed, 2007). Regardless of the impacts on scientists in developed nations, increased accessibility through open access does yield benefits to scientists from developing nations (Davis & Walters, 2011; J. A. Evans & Reimer, 2009).

The promise of open access for disseminating scientific information to the world at large has gained much less attention (Davis & Walters, 2011; Trench, 2008; for an exception see Heilman & West, 2015). Yet, more and more of the world turns to the Web for scientific information. For instance, as early as 1999 a full 20% of American adults sought medical and science information online (Miller, 2001).

What's more, one who actively seeks such information within the academic literature will quickly discover that, despite the paywalls, many important and impactful research articles are made freely available by their authors or third parties (Björk, Laakso, Welling, & Paetau, 2014; Wren, 2005). This is to say nothing of the fact that science may also be disseminated through distillation of its findings into venues like Wikipedia or science-centric websites and blogs so that, here too, the impact of open access may be limited. While full texts of the most impactful literature are, at least nominally, behind a paywall (Björk & Solomon, 2012), do Wikipedia's editors consult these texts? If they cite them in Wikipedia, have they consulted the full texts beyond a freely available abstract before referencing? If the academic literature is any guide, referenced material is sometimes consulted rather carelessly (Broadus, 1983; Rekdal, 2014). In short, the current understanding of the relationship between open access and the general public in the literature is limited at best (Davis & Walters, 2011).

### *Shortcomings and our contribution*

In addition to the role of accessibility, a number of substantive and methodological shortcomings remain. First, it is unclear if professional scientists edit Wikipedia's science articles. As we will show below, a preponderance of paywall references would suggest, albeit indirectly, this to be the case[10]. The scant existing evidence indicates that science articles are edited by people with general expertise, relative to the more narrow experts of popular culture articles (West et al., 2012). Second, most previous studies have completely ignored the articles that are never referenced on Wikipedia, thus sampling on the dependent variable. The only notable exception, (P. Evans & Krauthammer, 2011), treated the unreferenced articles outside the main analytic framework. While the framework treated (referenced) articles or journals as the unit of analysis, the unreferenced articles and journals were treated as a homogeneous group.

This study extends existing work in three chief ways. First, it models the role of accessibility (open access status) on referencing. Second, it covers *all* major research areas of science by observing rates at which Wikipedia references nearly 5,000 journals, accounting for nearly 20 million articles. Third, it treats unreferenced articles in the same analytic framework as those referenced. Yet the study is not without its own limitations, which are outlined more fully in the discussion section. Chief among these are that article-level characteristics are operationalized by the characteristics of the publishing journal. For example, the accessibility of articles is operationalized by their journal's open access policy, when, in fact, free access to many paywall articles exists through sanctioned or unsanctioned file-sharing (Björk et al., 2014; Wren, 2005). Thus, any observed advantage of open access referencing may be biased downward, i.e. an underestimate of the true effect (see the Conclusion for a discussion of measurement error).

## Data and Methods

### *Data sample*

### Journal data

Our analysis uses journal-level data from thousands of journals indexed by *Scopus*. Indexing over 21,000 peer-reviewed journals and with more than 2,800 classified as open access, *Scopus* is the world's largest database of scientific literature[11]. We obtained information on the 250 highest-impact journals within each of the following 26 major subject areas[12]: *Agricultural Sciences, Arts and Humanities, Biochemistry and General Microbiology, Business Management and Accounting, Chemical Engineering, Chemistry, Computer Science, Decision Sciences, Earth and Planetary Sciences, Economics and Finance, Energy Sciences, Engineering, Environmental Sciences, Immunology and Microbiology, Materials Sciences, Mathematics, Medicine, Neurosciences, Nursing, Pharmacology, Physics, Psychology, Social Science, Veterinary Science, Dental, Health Professions.* Assignment of journals to these broad subject areas is not exclusive; many journals fall into more than one category. As a result of cross listing, the list of candidate journals was less than 6500. The final data consisted of 4721 unique journals, 335 (7.1%) of which are categorized by the Directory of Open Access Journals as "open access."

Journals were also categorized more narrowly using the more than 300 "All Science Journal Classification" (ASJC) subject codes[13], e.g. Animal Science and Zoology, Biophysics, etc. These narrow codes were used to identify journals that address similar topics and thus indicate whether the journal is at risk for reference *vis-à-vis* demonstrated demand. Journals with at least one narrow subject code in common were considered "neighbors" and if at least one of these neighboring journals has been referenced the original "ego" journal was considered to be at risk for reference as well. Journals with no demonstrable demand were excluded from analysis. As an example, consider the journal *Science.* It is listed under (ASJC) subject code 1000 – general science. Other journals with this code – the "neighbors" of *Science* – are *Nature, PNAS*, and *Language Awareness*. *Language Awareness* is cross-listed under 5 others subject codes.

Impact factor was measured by the 2013 SCImago Journal Rank (SJR) impact factor. SJR correlates highly with the more conventional impact factor but takes into account self-citations and the diverse prestige of citing journals (González-Pereira, Guerrero-Bote, & Moya-Anegón, 2010; Leydesdorff, 2009). Table 1 displays the 15 highest SCImago impact journals, calculated with citations data available up to 2013.

| **Journal** | **Impact factor (SCImago2013)** |
|---|---|
| CA - A Cancer Journal for Clinicians | 45.894 |
| Reviews of Modern Physics | 34.830 |
| Annual Review of Immunology | 32.612 |
| Cell | 28.272 |
| Annual Review of Biochemistry | 27.902 |
| Quarterly Journal of Economics | 25.168 |
| Nature Genetics | 24.052 |
| Nature Reviews Genetics | 23.813 |
| Nature Reviews Molecular Cell Biology | 23.593 |
| Chemical Reviews | 23.543 |
| Nature | 21.323 |
| Acta Crystallographica Section D: Biological Crystallography | 20.717 |
| Advances in Physics | 20.349 |
| Annual Review of Cell and Developmental Biology | 19.686 |
| Annual Review of Neuroscience | 19.662 |

***Table 1.*** *15 highest-impact journals within* Scopus *according to* SCImago *impact factor (2013).*

## English Wikipedia data

References in the English Wikipedia were extracted from the 2014-11-15 database dump of all articles. We parsed every page and following (Nielsen, 2007) extracted all references that use Wikipedia's *cite journal* template. Since it allows editors to easily include inline references that are automatically rendered into an end-of-article bibliography, this template is the recommended way for editors to reference scientific sources in Wikipedia[14]. In all, there were 311,947 "cite-journal" tags in the English Wikipedia. An exploratory analysis of the 49 largest non-English Wikipedias can be found in the Appdenix.

### *Matching Scopus journals to Wikipedia references*

We checked each of the referenced journal names on Wikipedia against a list of *Scopus*-indexed journal names and common *ISI* journal name abbreviations. Of the 311,947 *cite-journal* tags, 203,536 could be linked to journals indexed by *Scopus*. Many of these references were non-unique, whereas our outcome of interest is whether articles from a journal are referenced on Wikipedia at all, not how many times. Therefore, to ensure that the counts for each journal included only unique articles, we distinguished articles by their DOIs and, if an article's DOI was not available, we used the article's title. Scopus' *coverage* of the output of various journals varies widely; our counts included only those articles published within the years of Scopus coverage.

In the end we matched 32,361 unique articles (and 55,262 total references) to our subset of *Scopus* journals (top 250 in each research field). 2,005 of the top *Scopus* journals are never referenced on the English Wikipedia. In most cases observed "journal names" that did not match to journals in *Scopus* were not academic journals but popular newspapers and magazines. Table A1 in the Appendix displays the 20 most frequently referenced sources that we were unable to link to *Scopus*. The top 3 non-science sources are *Billboard, National Park Service*, and *Royal Gazette*. However, efforts to match Wikipedia references to *Scopus* were imperfect, and the list also includes a handful of academic journals, including *The Lancet*.

### *Journal vs. article level unit of analysis*

We chose to take journals instead of individual articles as our unit of analysis for several reasons. First and most important, accessibility of articles, the focal point of this study, was measured at the journal level by whether the journal is or is not open access. Switching the unit of analysis to individual articles would have simply assigned the same value of accessibility to all articles from a particular journal. Second, while article-level citations are an attractive, finely grained metric, a journal's impact factor is also designed to capture citation impact, albeit more coarsely. The general topic of any given article is also well captured by the host journal's *Scopus*-assigned topic(s). Lastly, the matching of Wikipedia journal title strings to *Scopus* required some manual matching and these efforts were more practical at the level of thousands of journals instead of hundreds of thousands of articles.

### *percent_cited and Other Variables*

We present some of our results in terms of *percent_cited* -- the *percent of a journal's articles that are referenced on Wikipedia*. An equivalent interpretation of this journal-level metric is the *probability that a given article from a journal is referenced on Wikipedia.* Figure 2 illustrates the distribution (kde) of *percent_cited*.

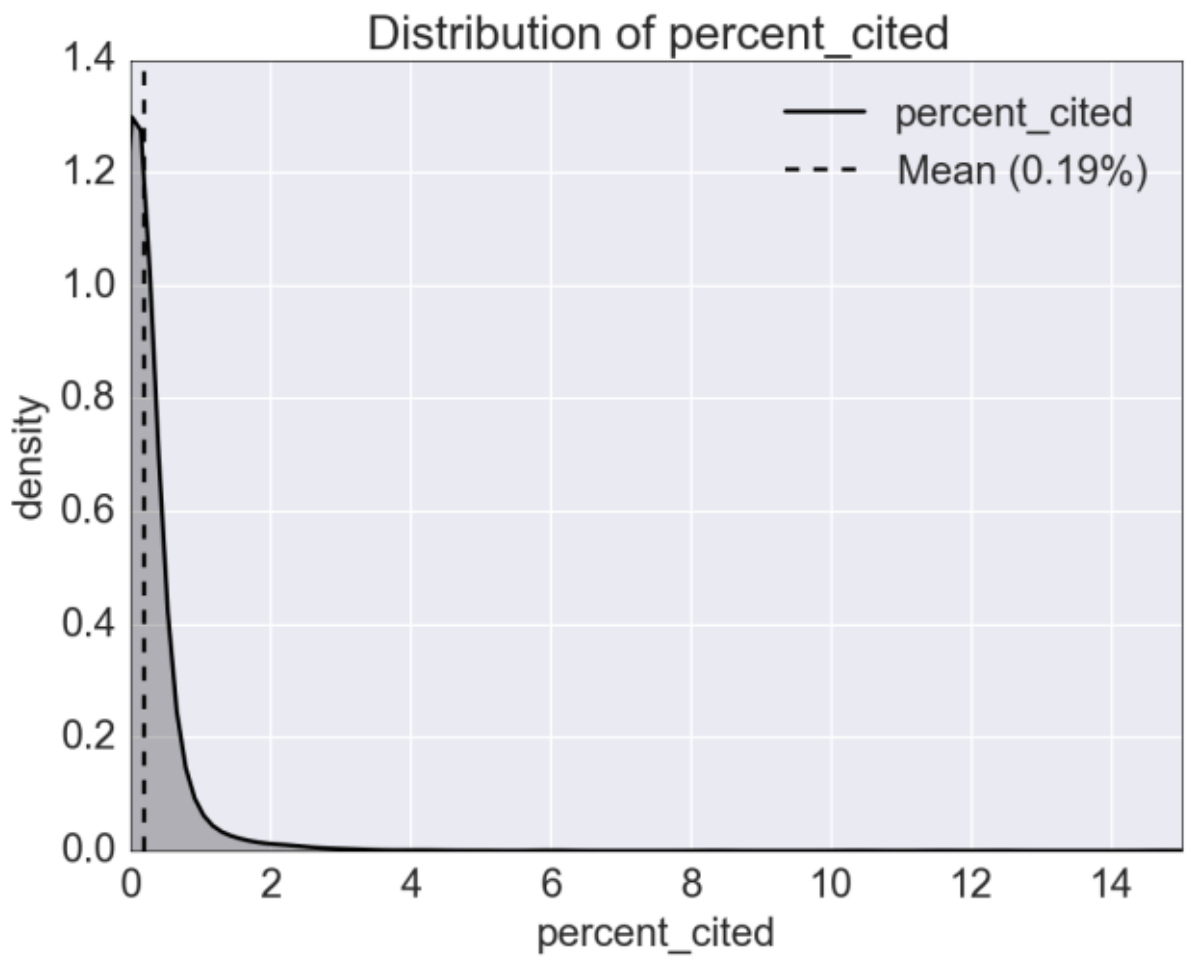

***Figure 2***. *Distribution (kde) of percent_cited of 4774 journals.*

As figure 2 demonstrates, the vast majority of journals that scientists use are referenced on the English Wikipedia very little: on average 0.19% of a journal's articles are referenced[15].

As mentioned above, the academic status of journals was measured using (SCImago) impact factors. To limit the influence of the few journals with uncommonly high impact factors the impact factor variable was (natural) log-transformed when used in the models. Figure 3 displays the distribution of impact factor and log-impact factor; to aid visualization only journals with impact factor <=15 are shown.

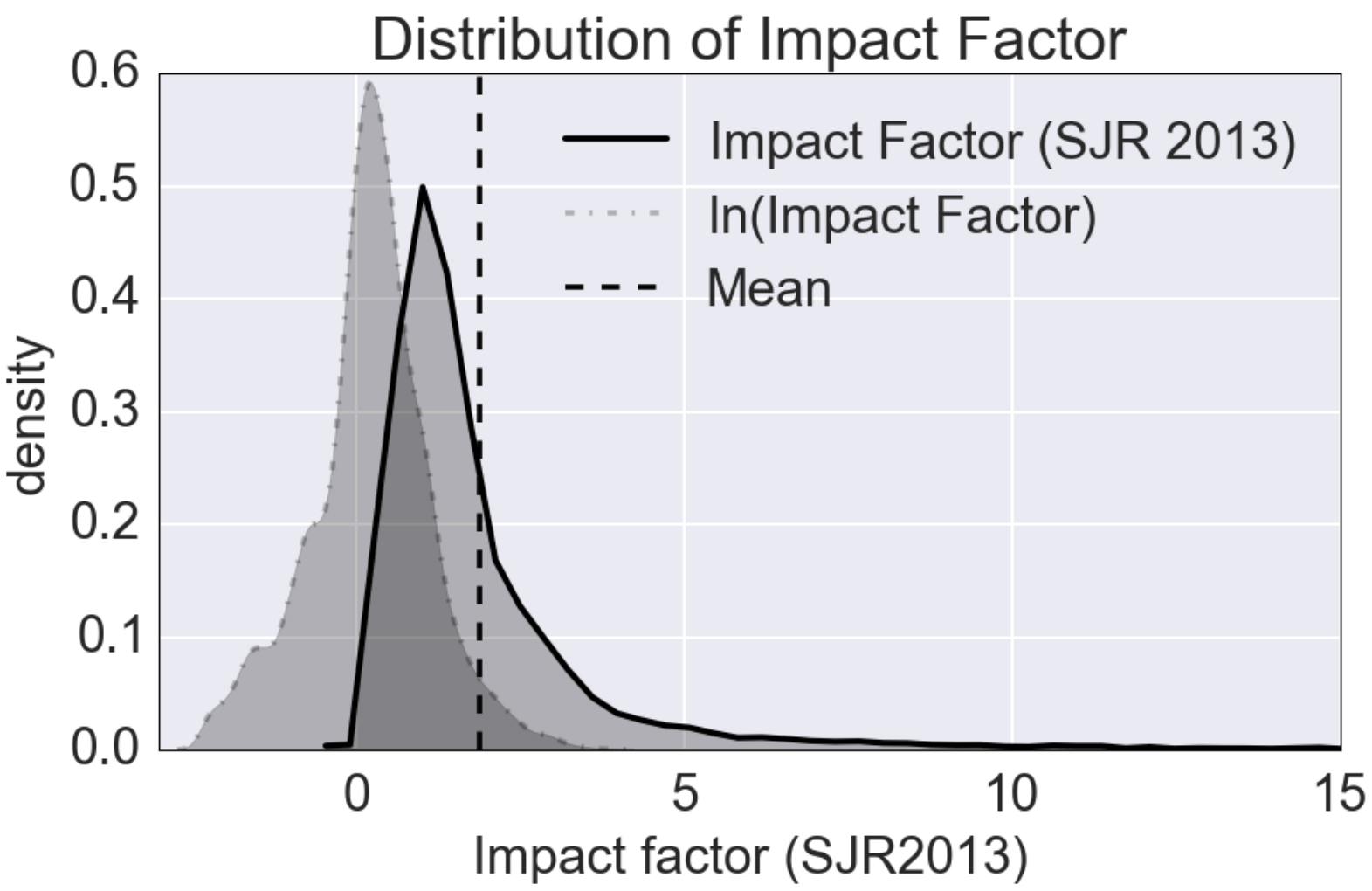

***Figure 3***. *Distribution (kde) of impact factor and ln(impact factor). To aid visualization, impact factor > 15 is not displayed.*

Table 2 presents the summary statistics for key variables: percent_cited, impact factor, ln(impact factor), and open access. Additionally, analyses use dummies for the 26 subject categories, e.g. psychology 0 or 1).

| ***Variable name*** | ***Mean*** | ***Std.*** | ***Min*** | ***Max*** |
|---|---|---|---|---|
| **percent_cited** | 0.193% | 0.545 | 0% | 14.7% |
| **impact factor** | 1.89 | 2.47 | 0.100 | 45.9 |
| **ln(impact factor)** | 0.212 | 0.909 | -2.30 | 3.83 |
| **open access** | 7.1% O.A. | ---- | 0 | 1 |

***Table 2****. Descriptive statistics of key variables.*

Lastly, Figure 4 displays a scatter plot of the key dependent variable, *percent_cited*, versus impact factor and open access.

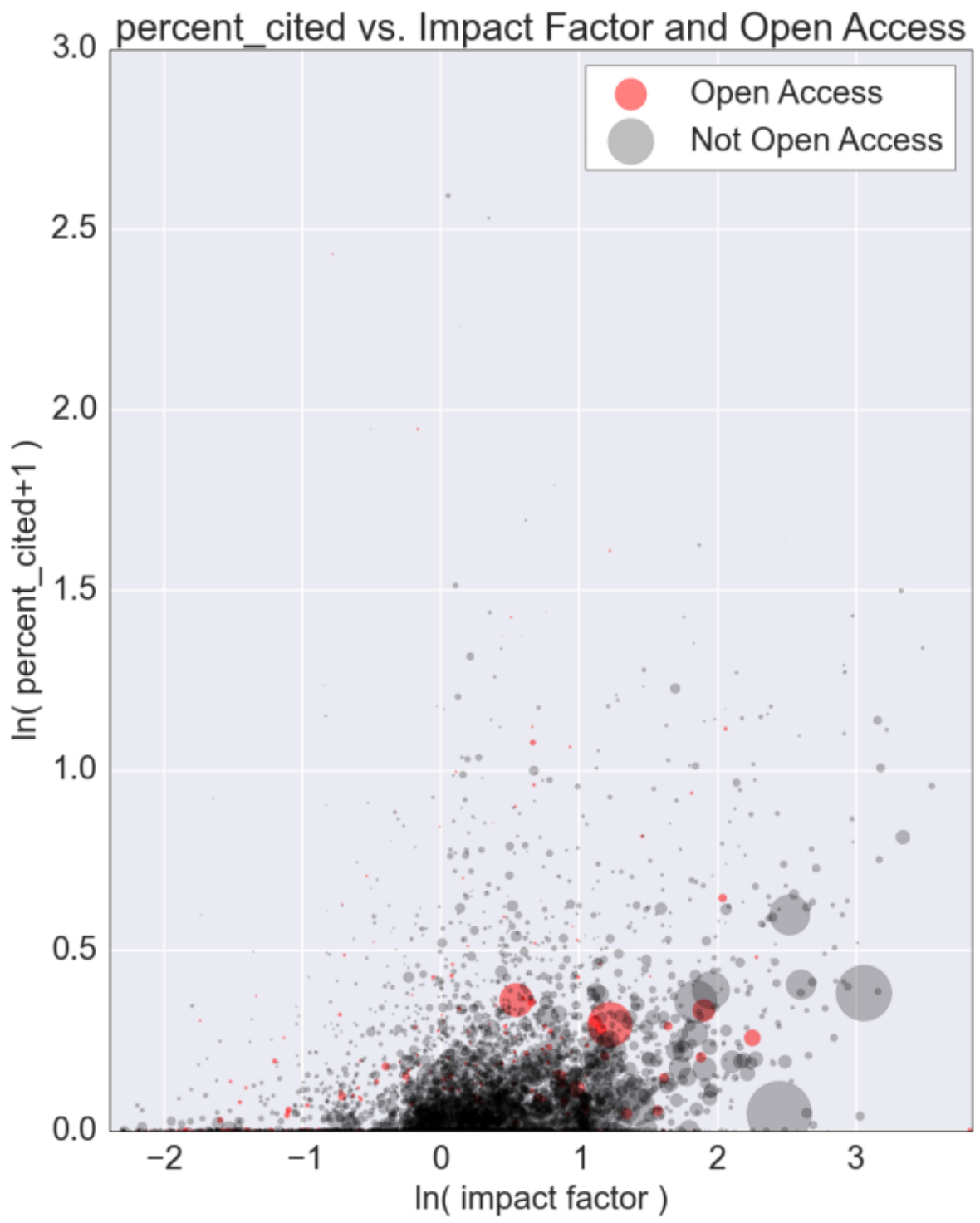

***Figure 4****. Scatter plot of journals' percent_cited vs. impact factor and open access policy. Dots are scaled by the total amount of articles published by each journal (and indexed by Scopus). Open access journals are shown with red dots.*

The scatter plots appear to show a modest relationship between a journal's impact factor and *percent_cited*, the percent of its articles referenced on Wikipedia, especially when considering journal size (dot size). The next section analyzes these relationships statistically.

## Results

We first present results of English Wikipedia's coverage. We ask, does Wikipedia draw equally on all branches of science? Next we focus on the role played by a journal's status and accessibility in predicting Wikipedia references. An exploratory analysis of references in the 49 largest *non-English* Wikipedias can be found in the Appendix.

## Coverage

Figure 5 below summarizes which branches of the scientific literature the English Wikipedia draws upon. The left panel shows the number of articles published by the top 250 journals in each field. The right panel shows the percent of those articles that are referenced at least once in the English Wikipedia.

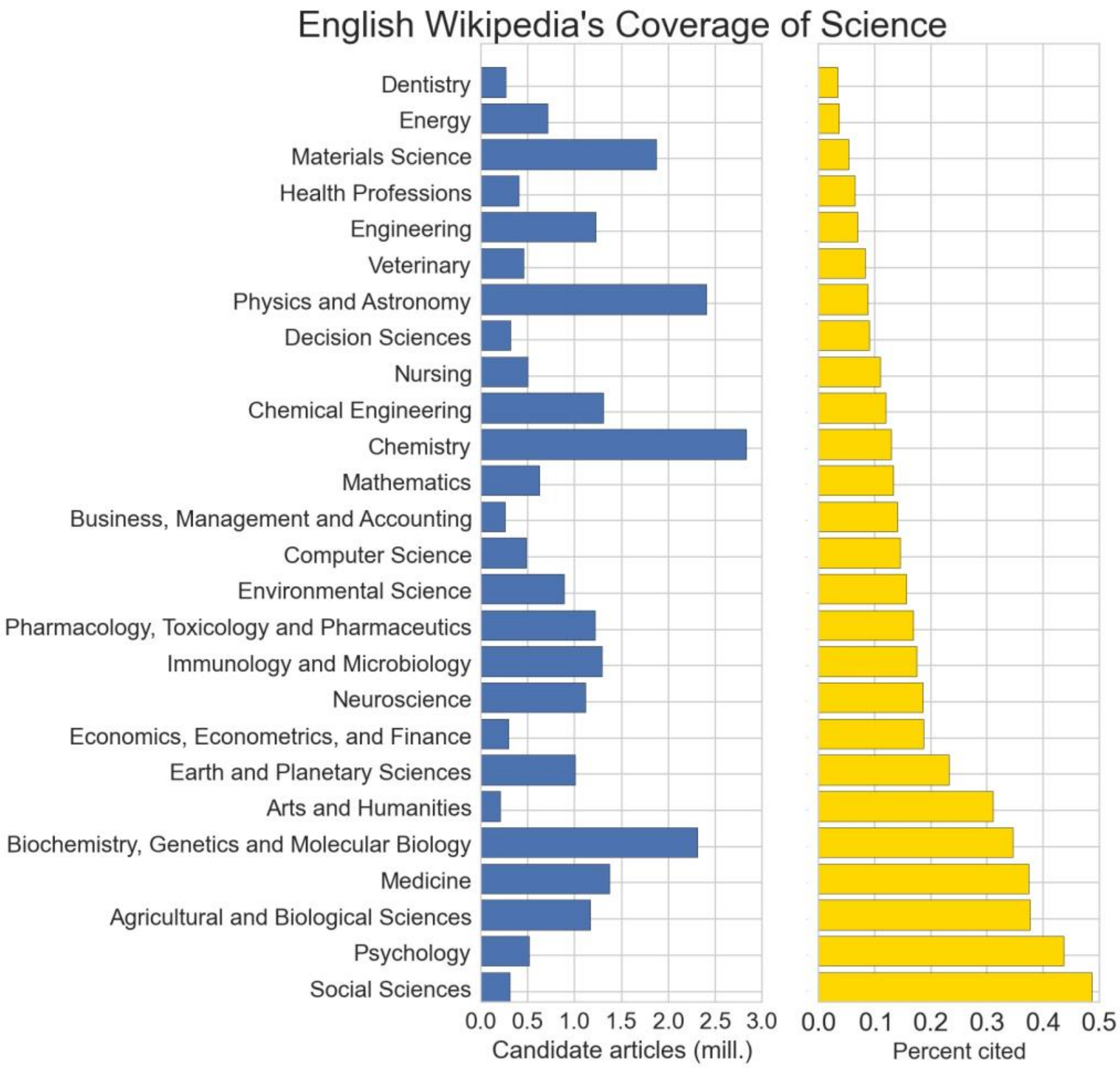

***Figure 5***. *English-language Wikipedia's coverage of academic research. Each field's candidate literature (left) consists of the articles published by the 250-highest impact journals within the field and indexed by Scopus. The right panel shows the percent of these articles referenced on Wikipedia.*

Figure 5 indicates that the coverage of science, as measured by the use of references, is very uneven and limited across scientific fields (Samoilenko & Yasseri, 2014). The social sciences represent a relatively small candidate literature but a relatively large portion of this literature is referenced on the English Wikipedia (0.4 – 0.5%). At the other end of the spectrum, dentistry, also a relatively small literature, is rarely referenced (< 0.05%). The ordering of disciplines by *percent_cited* does not engender a simple explanation. For example, such an ordering does not appear correlated with traditional distinctions like hard vs. soft science, or basic vs. applied. This finding is echoed by Nielsen (2007), who found that "computer and Internet–related journals do not get as many [references] as one would expect if Wikipedia showed bias towards fields for the 'Internet–savvy'". The highly uneven referencing across disciplines suggests that discipline should be controlled for in any statistical model, as is done below.

## Status and accessibility

We now present results from an intuitive statistical model that predicts the probability $p$ that an article from a given journal will be referenced given that journal's characteristics. The data-generating process is assumed to be a binomial process: each journal $i$ publishes $n_i$ articles and each of these articles is at risk $p_i$ of being referenced in Wikipedia, where $p_i$ depends on the journal. The probability that a journal $i$ has $k$ of its $n_i$ articles referenced in Wikipedia is thus

$$Pr(y_i = k \mid n_i, p_i) \sim \binom{n_i}{k} p_i^k (1 - p_i)^{n_i - k}$$

. $p_i$ is assumed to be a (logit) function of the journal characteristics $\mathbf{x_i}$'s (e.g. impact factor): $ln(\frac{p}{1-p}) = \mathbf{x}\beta$, where $\beta$ are the parameters to be estimated. The model just described is commonly used for proportional outcomes: it embeds the familiar logistic re-gression within a binomial process. This model is known as a generalized linear model (GLM) of the binomial-logit family (Hardin & Hilbe, 2012: 153-4)

Table 3 below displays estimates from this model of how journal characteristics are related to its *p*, probability of referencing, fitted to the English Wikipedia.

| **Variable** | **Odds ratio** | **95% C.I.** | **P-value** |
|---|---|---|---|
| **open_access** | 1.471 | 1.406, 1.539 | **0.000** |
| **log_sjr2013** | 1.879 | 1.852, 1.906 | **0.000** |
| **ag_bio_sciences** | 2.292 | 2.210, 2.377 | **0.000** |
| **arts_hum** | 1.836 | 1.689, 1.996 | **0.000** |
| **biochem_gen_mbio** | 1.059 | 1.030, 1.090 | **0.000** |
| **bus_man_acct** | 0.714 | 0.638, 0.799 | **0.000** |
| chem | 1.004 | 0.962, 1.048 | 0.863 |
| cheme | 0.968 | 0.912, 1.027 | 0.282 |
| cs | 0.991 | 0.916, 1.074 | 0.831 |
| decision_sci | 0.957 | 0.844, 1.084 | 0.489 |
| **dental** | 0.520 | 0.422, 0.614 | **0.000** |
| **earth_plan_sci** | 1.515 | 1.446, 1.587 | **0.000** |
| **econ_fin** | 1.106 | 1.010, 1.210 | **0.030** |
| **energy** | 0.551 | 0.487, 0.642 | **0.000** |
| **engineering** | 0.507 | 0.471, 0.545 | **0.000** |
| **envi_sci** | 0.743 | 0.703, 0.787 | **0.000** |
| **healthpro** | 0.787 | 0.696, 0.891 | **0.000** |
| **immu_micro_bio** | 1.114 | 1.065, 1.165 | **0.000** |
| **materials** | 0.640 | 0.598, 0.685 | **0.000** |
| **math** | 0.716 | 0.664, 0.772 | **0.000** |
| **medicine** | 0.660 | 0.642, 0.679 | **0.000** |
| neuro | 1.033 | 0.986, 1.081 | 0.168 |
| **nursing** | 1.206 | 1.101, 1.313 | **0.000** |
| **pharm** | 1.481 | 1.409, 1.556 | **0.000** |
| **phys** | 0.629 | 0.599, 0.660 | **0.000** |
| **psyc** | 2.628 | 2.504, 2.760 | **0.000** |
| **socialsci** | 1.357 | 1.283, 1.436 | **0.000** |
| **vet** | 0.898 | 0.807, 1.000 | **0.048** |

***Table 3.*** *Estimates from the GLM estimated on English Wikipedia reference data. Variables with statistically* in*significant odds ratios are not bolded. n= 4720, df=28*

The column of odds ratios indicates how the odds of referencing change with unit changes in the independent variables. For indicator variables, e.g. open access, these ratios are interpreted as the increase in odds when the indicator is true. For example, the odds that an article is referenced on Wikipedia increase by 47% if the article is published in an open access journal.

To interpret these results in terms of probabilities rather than odds ratios we must evaluate the model at particular values of the variables. Figure 6 displays the observed and predicted references for a range of values of impact factor and *open_access*. The indicator variables designating particular disciplines are set at their modes (0).

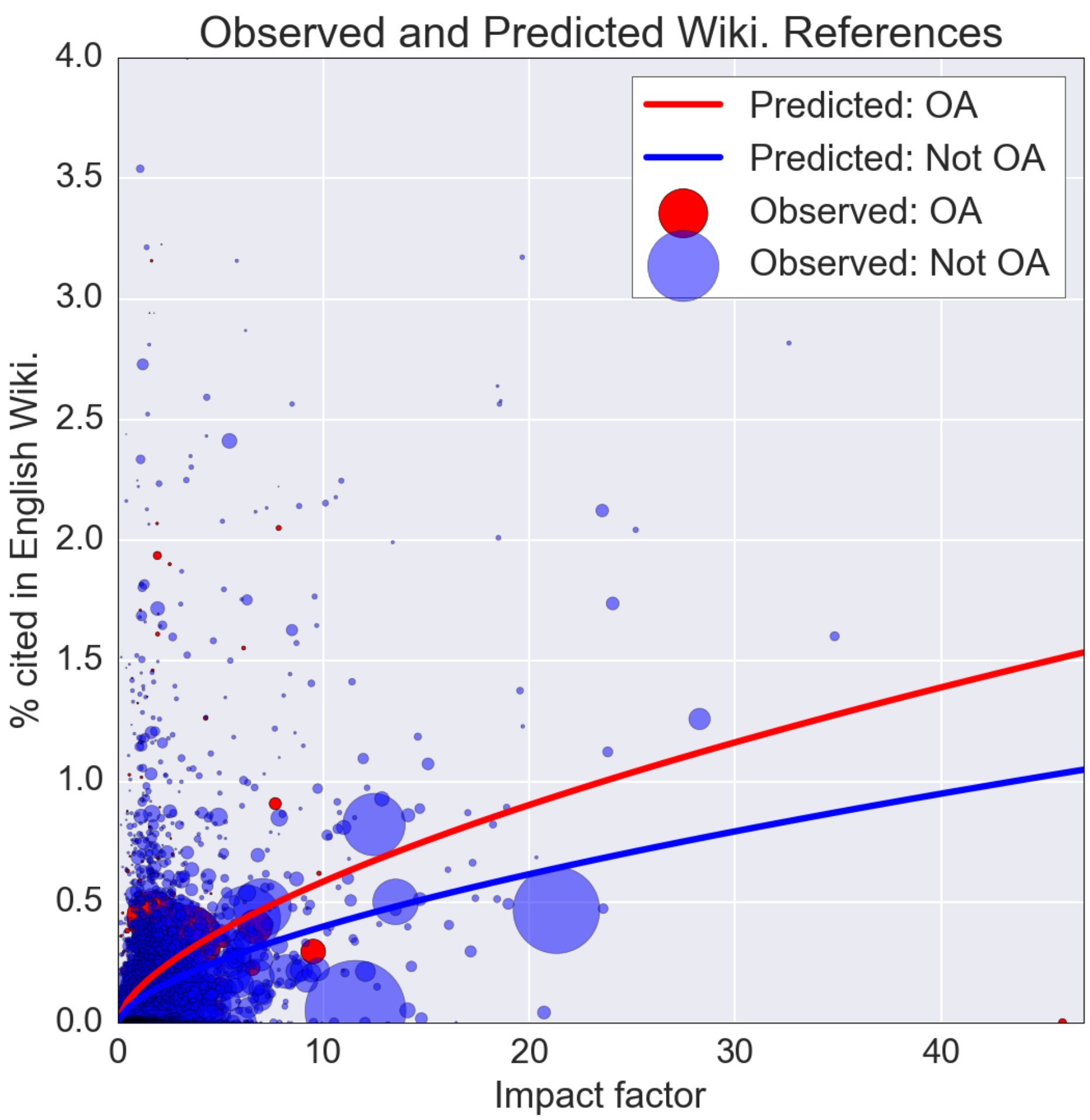

***Figure 6.*** *Observed (dots) and predicted (solid lines) English Wikipedia references. Red dots designate open access journals. The dot size is proportional to the number of articles the journal published.*

The figure demonstrates that a journal's impact factor has positive and asymptotic effect on the percent of its contents referenced in the English Wikipedia (*percent_cited*). Open access journals (red dots) are relatively uncommon, but these journals are referenced more often than paywall journals of similar impact factor. For example, in our sample of psychology journals, open access journals have an average impact factor of 1.59, while closed access journals have an average impact factor of 1.77. Yet in the English Wikipedia, editors reference an average of 0.49% of open access journals' articles but only 0.35% of closed access journals' articles, despite the higher impact factors.

## Conclusion

This article examined in unprecedented detail and scale how the English language Wikipedia references the scientific literature. Of central interest was the relationship between an articles' academic status and accessibility on its probability of being referenced in Wikipedia. In the appendix, we make a cursory attempt to extend this analysis to the world's 50 largest Wikipedias. Previous studies have focused only on the role of academic status on referencing in the English Wikipedia and have often ignored unreferenced articles. In contrast, we began by identifying an enormous (~19.4MM articles) corpus of scientific literature that scientists routinely cite, found a subset of this literature for which Wikipedia editors demonstrate demand, and estimated a statistical model to identify the features of journals that predict referencing.

We found that a journal's academic status (impact factor) routinely predicts its appearance on Wikipedia. We also demonstrated, for the first time, that a journal's accessibility (open access policy) generally increases probability of referencing on Wikipedia as well, albeit less consistently than its impact factor. The odds that an open access journal is referenced on the English Wikipedia are about 47% higher compared to closed access, paywall journals. Moreover, of closed access journals, those with high impact factors are also significantly more likely to appear in the English Wikipedia. Therefore, editors of the English Wikipedia act as "distillers" of high quality science by interpreting and distributing otherwise closed access knowledge to a broad public audience, free of charge. Moreover, the English Wikipedia, as a platform, acts as an “amplifier” for the (already freely available) open access literature by preferentially broadcasting its findings to millions.

### *Limitations and directions for future research*

Our findings are not without limitations. First and foremost, it bears emphasis that this study did not investigate the nature of Wikipedia’s sources as a whole (see Ford et al., 2013 for an excellent examination of sources). Only a fraction of Wikipedia’s references use the scientific literature, and this is the subset on which we focused. Consequently the present study cannot address the concern expressed by others, e.g. (Luyt & Tan, 2010), that sources outside the scientific literature are used too heavily in scientific articles. Second, the study was cross-sectional in nature; it is conceivable that open access articles differ from closed access, paywall articles in their relevance to Wikipedia. Future work can test the potential confounding factor of unmeasured relevance by observing reference rates for articles which have been experimentally assigned to open and closed-access statuses, as has been done by some psychology journals (Davis et al., 2008).

Third, the study measured accessibility of articles by the open-access policy of the publishing journals. However, many articles in paywall journals are made freely available by their authors or third parties (Björk et al., 2014; Wren, 2005). The resulting error in the measurement of accessibility may bias the observed advantage of open access in either direction: if open access *articles* from paywall *journals*, erroneously coded as closed access, are referenced at higher rates than the journals’ truly closed access articles (Gargouri et al., 2010; Harnad & Brody, 2004), the *true* advantage of open access will be even higher than we observed. In the (unlikely) case that open access articles in paywall journals are referenced less than closed access articles, the observed open access advantage will be an overestimate. The academic status of articles is also operationalized by a journal characteristic – its impact factor. In fact, many articles out- or under-perform their journal’s impact factor. While this measurement error likely adds noise to the data, it probably does so without biasing the estimated effect of impact factor on referencing in one direction or another.

### *The impact of open access science*

The chief finding of this study bears emphasis. We believe the existing discussion of open access has focused too narrowly on the academic literature. Early results showing that open access improves scientific outcomes such as citations have been tempered by newer experimental evidence showing small to null causal effects, and a lively debate has ensued. Our research shifts focus to diffusion, showing that open access policies have a tremendous impact on the diffusion of science to the broader general public through an intermediary like Wikipedia. This effect, previously a matter primarily of speculation, has empirical support. As millions of people search for scientific information on Wikipedia, the science they find distilled and referenced in its articles consists of a disproportionate quantity of open access science.

## Notes

1. http://www.alexa.com/siteinfo/wikipedia.org
2. https://en.wikipedia.org/wiki/Wikipedia:Verifiability. Accessed 2015-06-15.
3. http://stats.wikimedia.org/EN/TablesPageViewsMonthly.htm. Accessed 2015-06-16.
4. Wikipedia has recently partnered with major publishers to provide editors access to some paywall literature: http://en.wikipedia.org/wiki/Wikipedia:The_Wikipedia_Library. Accessed 2015-09-02.
5. http://en.wikipedia.org/wiki/Wikipedia:Notability. Accessed 2015-06-11.
6. http://en.wikipedia.org/wiki/Wikipedia:Five_pillars. Accessed 2015-05-29.
7. http://en.wikipedia.org/wiki/Wikipedia:Identifying_reliable_sources#Scholarship. Accessed 2015-05-29.
8. Citation metrics often influence the ranking of academic search results and may thus promote high impact journals without searchers' knowledge.
9. This literature has grown to thousands of items and is impossible to summarize fully. See (Craig, Plume, McVeigh, Pringle, & Amin, 2007; Davis & Walters, 2011) for two reviews of parts of this literature.
10. As corroborating evidence consider the list of Wikipedia editors by [self-reported] degree lists more than 1000 users with PhDs: http://en.wikipedia.org/wiki/Category:Wikipedians_with_PhD_degrees. Accessed 2015-09-02.
11. http://www.elsevier.com/online-tools/scopus/content-overview. Accessed 2015-01-24.
12. The subject area "general" was excluded because it contained only four journals, all of which were cross-listed with other top-level categories.
13. http://info.sciencedirect.com/scopus/scopus-in-detail/content-coverage-guide/journalclassification. Accessed 2015-06-03.

14. Editors may also reference articles in other ways, for example by providing in-line links. We focus on the "cite-journal" template for three reasons. First, it shows clear intent to reference. Second, it has been used in previous research. Lastly, Ford and colleagues (Ford, Sen, Musicant, & Miller, 2013) found that "<ref>" tags were used most often to reference sources, and the "cite-journal" templates on which we focus are nested within such <ref> tags.
15. 2,005 (out of 4,721) journals are never referenced at all (*percent_cited* = 0).

## Acknowledgements

This research was enabled by grant 39147 to the Metaknowledge Network by the John Templeton Foundation. An earlier version of this work was presented at the Wikipedia Workshop of the 9th International Conference on Web and Social Media, Oxford, UK. We thank the reviewers for perceptive comments that greatly improved this article.

## Appendix

**Table A1.** Most common sources referenced using the *cite journal* template that are not indexed by *Scopus*.

| Journal name | Times referenced |
|---|---|
| BILLBOARD | 1630 |
| NATIONAL PARK SERVICE | 539 |
| ROYAL GAZETTE | 523 |
| BULL AMER MATH SOC | 506 |
| FLIGHT INTERNATIONAL | 455 |
| BAH NEWS | 385 |
| NEW YORK TIMES | 369 |
| ROLLING STONE | 360 |
| ENTERTAINMENT WEEKLY | 342 |
| J BOMBAY NAT HIST SOC | 314 |
| WHOS WHO | 312 |
| BIZJOURNALSCOM | 288 |
| THE GUARDIAN | 287 |
| THE LANCET | 281 |
| VARIETY MAGAZINE | 270 |
| INSIDE SOAP | 239 |
| TIME MAGAZINE | 238 |
| BIOCHIMICA ET BIOPHYSICA ACTA | 201 |
| EDMONTON JOURNAL | 201 |
| SPORTS ILLUSTRATED | 197 |

## Non-English Wikipedias

Non-English Wikipedias have been noticeably neglected by the research community (Mesgari, Okoli, Mehdi, Nielsen, & Lanamäki, 2015; Schroeder & Taylor, 2015). It is thus important to test whether any of the findings of this article extend to the millions of articles in non-English Wikipedias. Below we present an exploratory analysis of scientific references in the 49 largest non-English Wikipedias.

### Data

Database dumps of the 49 largest non-English Wikipedias were downloaded on 2015-05-10. For each of these, we extracted tags containing “journal” or “doi”. Thus the process for obtaining scientific references in non-English Wikipedias did not take into account language-specific tags. Non-English Wikipe-dias may also reference domestic scientific journals that are not indexed by *Scopus*. Thus, this exploratory approach surely undercounts scientific references to non-English Wikipedias.

The English Wikipedia referenced by far the greatest number of unique articles. Figure A.1 displays the number of unique articles referenced in other Wikipedias, sorted by size (total articles).

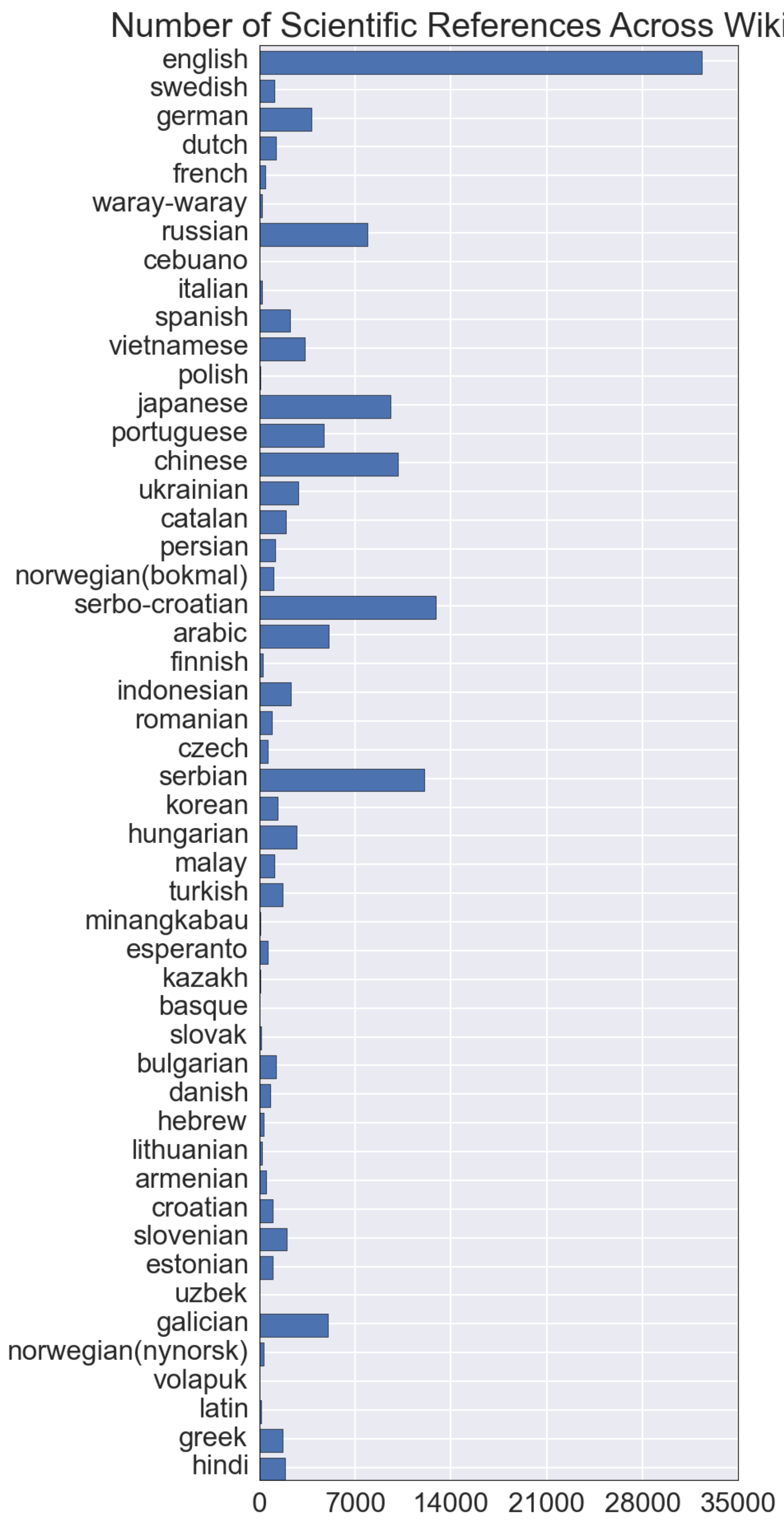

***Figure A.1****. Number of unique scientific articles referenced on the 50 largest Wikipedias.*

### *Empirical Strategy*

Certainly not all findings published in the academic literature belong on Wikipedia. Only small subsets of published findings are important and notable enough to be referenced in Wikipedia. Ideally, studies of how Wikipedia editors reference sources should explain which items in this smaller subset are and are not referenced. Nevertheless, previous studies have struggled to distinguish the candidate articles that are at risk for reference from those that do not belong on Wikipedia. Yet, to model referencing decisions with *all* articles – including the dozens of millions of articles never referenced on Wikipedia – is likely to result in a model that predicts that no article will ever be referenced. Consequently most studies have voluntarily hobbled themselves by simply modeling only on the subset of referenced articles.

Here we propose a compromise strategy based on "demonstrated demand." The idea is simple: articles are at risk for reference if *other* articles on the same topic are referenced. Topical reference indicates that there is demand from Wikipedia editors for literature on the topic and that an article's characteristics (e.g. accessibility, status) may determine which of the candidate articles an editor finds and references. Conversely, if articles on a given topic are never referenced, it is likely that Wikipedia editors do not "demand" literature on this topic, no matter the accessibility or status of the supply. Demonstrated demand exists at the level of topics and, like accessibility and status, we identify an article's topic at the journal level. Demonstrable demand is also a language-dependent metric: some Wikipedias may lack editors with expertise or interest in, for example, dentistry, thereby consigning all dentistry journals to irrelevance with regards to referencing decisions (but not irrelevant for analysis of coverage, of course). To calculate demonstrated demand we identify for each journal its topical "neighbors" and assign demand of 0 if none of these neighbors are ever referenced in a particular Wikipedia.

We calculate demonstrated demand for a journal through its topical neighbors, which are defined as other journals that share at least one narrow (ASJC) subject code. Only 1 journal, *Prevenzione & assist-* *enza dentale*, had no neighbors while the mean neighborhood size was 144.8. Figure A4 displays the distribution (kde) of neighborhood size.

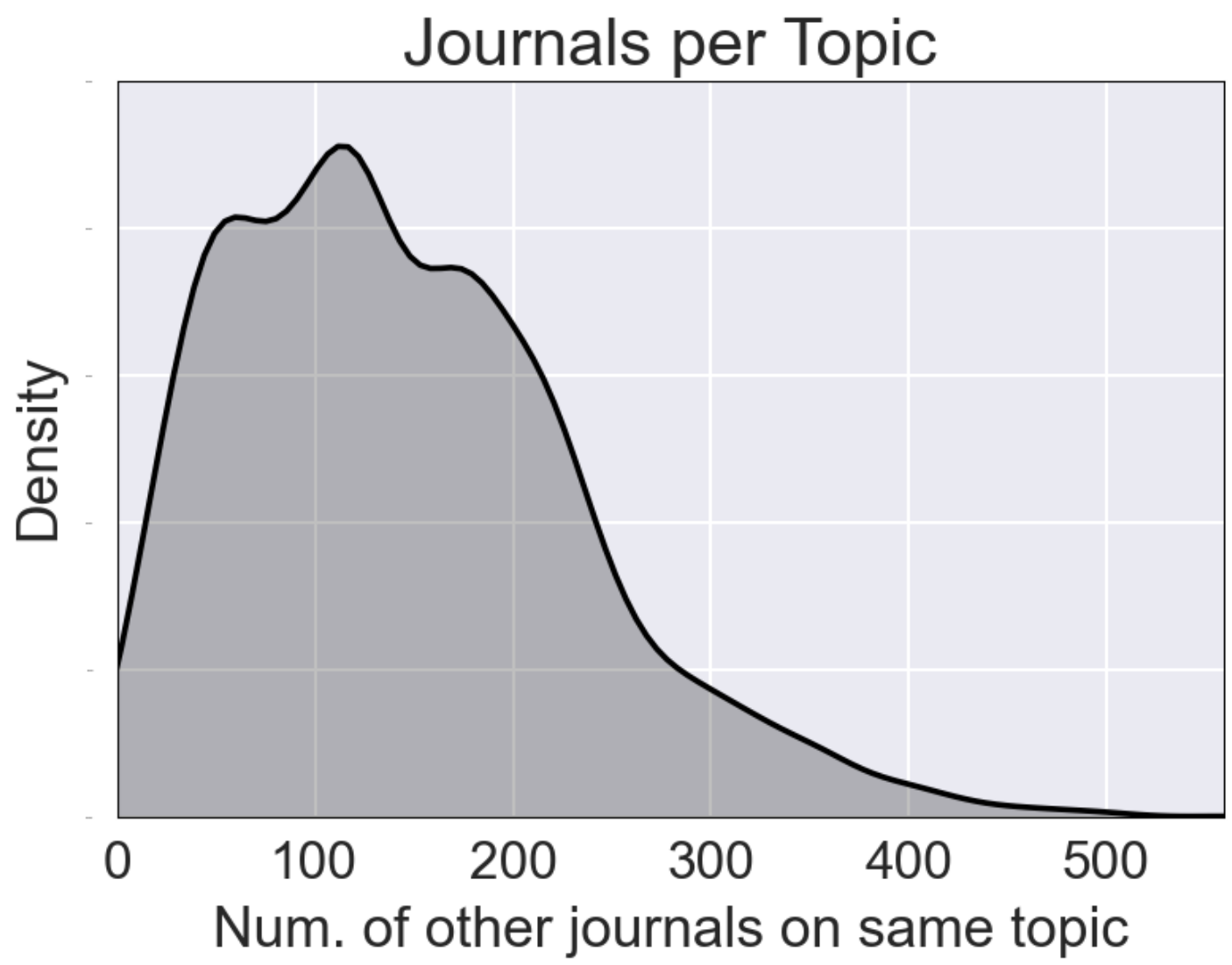

***Figure A4***. *Distribution of the topical neighborhood sizes of journals. On average journals had 144.8 other journals that addressed the same topic(s).*

Table A2 contains the percentages of journals that were excluded from estimating models in each language. This percentage varies widely. For example, only 0.17% of journals were not used for the English Wikipedia model, 49.87% for Slovak, and a 100% for Volapuk. These numbers correspond directly to demonstrable demand for various research literatures by the editors of each Wikipedia. While the English Wikipedia references ~32,000 articles from top journals, the Slovak Wikipedia references only 108 and Volapuk references 0.

**Table A2.** Percent of journal data that is not used in estimates language-specific models (demonstrated demand = 0). These percentages indicate the portion of research areas for which there is no demonstrable demand from (language-specific) Wikipedia editors.

| Wikipedia language | Percent excluded (weight=0) |
|---|---|
| **chinese** | 1.28 |
| **russian** | 1.28 |
| **japanese** | 1.36 |
| **arabic** | 1.47 |
| **vietnamese** | 2.01 |
| **portuguese** | 2.16 |
| **german** | 2.30 |
| **spanish** | 2.62 |
| **indonesian** | 2.95 |
| **hindi** | 3.62 |
| **hungarian** | 3.73 |
| **ukrainian** | 4.17 |
| **slovenian** | 4.25 |
| **persian** | 4.40 |
| **serbian** | 4.92 |
| **greek** | 4.99 |
| **turkish** | 5.63 |
| **serbo-croatian** | 6.35 |
| **malay** | 6.85 |
| **bulgarian** | 7.12 |
| **dutch** | 7.31 |
| **catalan** | 7.88 |
| **danish** | 8.36 |
| **swedish** | 9.78 |
| **romanian** | 10.08 |
| **korean** | 10.08 |
| **estonian** | 11.73 |
| **galician** | 12.13 |
| **norwegian(bokmal)** | 12.36 |
| **czech** | 12.84 |
| **french** | 13.68 |
| **croatian** | 14.52 |
| **hebrew** | 20.93 |
| **armenian** | 22.62 |
| **waray-waray** | 25.32 |
| **esperanto** | 26.37 |
| **finnish** | 26.81 |
| **italian** | 29.16 |
| **lithuanian** | 34.00 |
| **norwegian(nynorsk)** | 40.01 |
| **latin** | 43.67 |
| **slovak** | 49.87 |
| **polish** | 62.69 |
| **uzbek** | 78.91 |
| **minangkabau** | 80.88 |
| **kazakh** | 81.44 |
| **basque** | 86.30 |
| **cebuano** | 86.95 |
| **volapuk** | 99.98 |

## Results

From each Wikipedia's model, two parameters are of focal interest: the odds ratio (of probability of referencing) when open access is True, and odds ratio when (log of) impact factor increases by 1 unit. Table A3 shows these odds ratios and associated p-values for each Wikipedia. Ratios significant at the 0.05 level are bolded.

**Table A3.** Odds ratios and associated p-values for *open access* and (log) *impact factor* for 50 Wikipedias. Statistically significant odds ratios are bolded.

| Wikipedia Language | open access odds ratio | open access p-value | Ln(impact factor) odds ratio | Ln(impact factor) p-value |
|---|---|---|---|---|
| arabic | 0.923 | 0.258 | **2.189** | **0.000** |
| armenian | 1.052 | 0.802 | **2.669** | **0.000** |
| basque | 0.000 | 1.000 | 1.059 | 0.890 |
| bulgarian | 0.936 | 0.623 | **2.256** | **0.000** |
| catalan | **1.452** | **0.000** | **1.963** | **0.000** |
| cebuano | 0.000 | 1.000 | 1.989 | 0.093 |
| chinese | **1.337** | **0.000** | **2.257** | **0.000** |
| croatian | **0.651** | **0.009** | **2.230** | **0.000** |
| czech | 1.247 | 0.230 | **2.258** | **0.000** |
| danish | 0.722 | 0.120 | **2.190** | **0.000** |
| dutch | **1.743** | **0.000** | **2.238** | **0.000** |
| english | **1.471** | **0.000** | **1.878** | **0.000** |
| esperanto | 1.114 | 0.591 | **1.245** | **0.001** |
| estonian | 1.221 | 0.151 | **2.705** | **0.000** |
| finnish | 0.666 | 0.300 | **1.576** | **0.000** |
| french | 0.850 | 0.550 | **2.030** | **0.000** |
| galician | **1.464** | **0.000** | **2.176** | **0.000** |
| german | **1.755** | **0.000** | **2.264** | **0.000** |
| greek | 0.798 | 0.078 | **2.008** | **0.000** |
| hebrew | 1.191 | 0.531 | **1.906** | **0.000** |
| Hindi | **0.757** | **0.029** | **2.113** | **0.000** |
| hungarian | **0.749** | **0.005** | **1.804** | **0.000** |
| indonesian | 0.886 | 0.242 | **2.467** | **0.000** |
| italian | 0.638 | 0.299 | **2.072** | **0.000** |
| japanese | **2.577** | **0.000** | **1.865** | **0.000** |
| kazakh | 0.000 | 1.000 | 1.759 | 0.114 |
| **korean** | 1.246 | 0.055 | **1.944** | **0.000** |
| **latin** | 0.812 | 0.637 | **1.975** | **0.000** |
| **lithuanian** | 1.035 | 0.923 | **2.345** | **0.000** |
| **malay** | 1.211 | 0.157 | **2.214** | **0.000** |
| **minangkabau** | 2.927 | 0.314 | 1.661 | 0.224 |
| **norwegian(bokmal)** | 0.866 | 0.437 | **2.328** | **0.000** |
| **norwegian(nynorsk)** | 0.588 | 0.109 | **1.510** | **0.000** |
| **persian** | 0.941 | 0.678 | **2.210** | **0.000** |
| **polish** | 0.588 | 0.480 | **2.330** | **0.000** |
| **portuguese** | **1.527** | **0.000** | **2.076** | **0.000** |
| **romanian** | 0.903 | 0.545 | **2.178** | **0.000** |
| **russian** | **1.419** | **0.000** | **2.086** | **0.000** |
| **serbian** | **3.824** | **0.000** | **1.516** | **0.000** |
| **serbo-croatian** | **3.761** | **0.000** | **1.518** | **0.000** |
| **slovak** | 1.943 | 0.157 | **3.249** | **0.000** |
| **slovenian** | 0.926 | 0.487 | **2.389** | **0.000** |
| **spanish** | **1.913** | **0.000** | **1.698** | **0.000** |
| **swedish** | **3.745** | **0.000** | **2.094** | **0.000** |
| **turkish** | **1.262** | **0.021** | **2.956** | **0.000** |
| **ukrainian** | **0.818** | **0.030** | **2.566** | **0.000** |
| **uzbek** | 0.000 | 1.000 | **2.963** | **0.012** |
| **vietnamese** | 0.966 | 0.682 | **2.143** | **0.000** |
| **volapuk** | 0.588 | 0.480 | **2.330** | **0.000** |
| **waray-waray** | **2.104** | **0.017** | **2.172** | **0.000** |

While earlier results showed that both accessibility and status increase the odds that a journal will be referenced in the English Wikipedia, the relative strength of these effects varies across languages. Some Wikipedias, like the Turkish, prioritize a journal's academic status over accessibility; the odds of referencing high status journals are nearly 200% higher than lower status journals. Other Wikipedias, like the Serbian, prioritize accessibility over status; the odds of referencing an open access journals are ~275% higher than a paywall journal.

Intuition and previous work suggests poorer countries rely on open access literature more (J. A. Evans & Reimer, 2009), yet this pattern is not apparent in Figure 8. For example, India and Ukraine, relatively poor countries naturally associated with the Hindi and Ukrainian Wikipedias, actually exhibit a small preference against open access literature, while a relatively wealthy country like Sweden has a Wikipedia that exhibits a huge preference for open access literature. The unexpected patterns may be due to the influence of bots (Steiner, 2014). For example, about a third of all articles on the Swedish Wikipedia were created by a bot (Jervell, 2014). Idiosyncrasies of the small number of human and non-human entities that edit science in non-English Wikipedias may thus play a larger role than gross cross-national patterns.

It bears emphasis that this analysis of references in non-English Wikipedias is exploratory. Further work should extract references in a way that is sensitive to each Wikipedia's language and conventions. Such analysis may reveal differences in how scientific content found in Wikipedia across languages is differentially embedded in or husbanded by local scientific communities.